\documentstyle[12pt]{article} 
\normalbaselines
\begin{document} 
\bibliographystyle{unsrt}

\begin{center} 
{\LARGE \bf Discrete Differential Geometry and \newline Lattice Field Theory}\\[5mm]

Miguel Lorente\\{\it Departamento de F\'{\i}sica, Universidad de Oviedo, 
33007 Oviedo, Spain}\\[5mm]  
\end{center}

\begin{abstract} 

We develope a difference calculus analogous to the differential geometry by translating the forms and exterior derivatives to similar
expressions with difference operators, and apply the results to fields theory on the lattice [Ref. 1]. Our approach has the advantage with
respect to other attempts [Ref. 2-6] that the Lorentz invariance is automatically preserved as it
can be seen explicitely in the Maxwell, Klein-Gordon and Dirac equations on the lattice.

\end{abstract}

\section{A difference calculus of several independent \newline variables}

Given a function of one independent variable the forward and
backward differences are defined as
\[ \Delta f(x) \equiv f(x + \Delta x) - f(x) \quad , \quad \nabla f(x) \equiv f(x) - f(x - \Delta x)\]

Similarly, we can define the forward and backward promediate operator
\[\widetilde{\Delta }f(x) \equiv {1 \over 2}\left\{{f(x+\Delta x) + f(x)}\right\} \quad , \quad
\widetilde{\nabla }f(x) \equiv {1 \over 2}\left\{{f(x-\Delta x) + f(x)}\right\}\]

Hence the difference or promediate of the product of two functions follows:
\begin{eqnarray}
\Delta \left\{{f(x) g(x)}\right\} &=& \Delta f(x) \widetilde{\Delta } g(x) + \widetilde{\Delta } f(x) \Delta g(x) \\
\widetilde{\Delta } \left\{{f(x) g(x)}\right\} &=& \widetilde{\Delta } f(x) \widetilde{\Delta } g(x) + {1 \over 4}\Delta f(x) \Delta g(x)
\end{eqnarray}

This calculus can be enlarged to functions of several independent variables. We use the following
definitions:
\begin{eqnarray*}
{\Delta }_{x} f(x,y) &\equiv & f(x+\Delta x,y) - f(x,y) \\
{\Delta }_{y} f(x,y) &\equiv & f(x,y+\Delta y) - f(x,y) \\
{\widetilde{\Delta }}_{x} f(x,y) &\equiv & \frac{1}{2} \left\{{f (x+\Delta x,y) + f(x,y)}\right\} \\
{\widetilde{\Delta }}_{y} f(x,y) &\equiv &{1 \over 2} \left\{{f (x,y+\Delta y) + f(x,y)}\right\} \\
\Delta f(x,y) &\equiv & f (x+\Delta x , y+\Delta y) - f(x,y) \\
\widetilde{\Delta } f(x,y) &\equiv & {1 \over 2} \left\{{f (x+\Delta x , y+\Delta y) + f(x,y)}\right\}
\end{eqnarray*}

These definitons can be easily generalized to more independent variables but for the sake of
brevity we restrict ourselves to two independent variables. From the last definitions it can be
proved the following identities: 
\begin{eqnarray}
\Delta f(x,y) &=& {\Delta }_{x} {\widetilde{\Delta }}_{y} f(x,y) + {\widetilde{\Delta }}_{x} {\Delta }_{y} f(x,y) \\
\widetilde{\Delta } f(x,y) &=& {\widetilde{\Delta }}_{x} {\widetilde{\Delta }}_{y} f(x,y) + {1 \over 4}{\Delta }_{x} {\Delta }_{y}f(x,y)
\end{eqnarray}

We can also construct the difference calculus for composite functions. For the sake of
simplicity we restrict ourselves to functions of two dependent variables and two independent
ones, $f \left({u (x,y) , v (x,y)}\right)$.

We define:
\begin{eqnarray*}
{\Delta }_{u} f &\equiv & f (u+\Delta u, v) - f (u,v) \\
{\Delta }_{v} f &\equiv & f (u,v+\Delta v) - f (u,v) \\
{\widetilde{\Delta }}_{u} f &\equiv & {1 \over 2} \left\{{f (u+\Delta u, v) + f (u,v)}\right\} \\
{\widetilde{\Delta }}_{v} f &\equiv & {1 \over 2} \left\{{f (u,v+\Delta v) + f (u,v)}\right\} \\
{\Delta }_{x} f &\equiv & {f \left({u (x+\Delta x, y), v (x+\Delta x, y)}\right) - f \left({u (x,y), v (x,y)}\right)} \\
{\Delta }_{y} f &\equiv & {f \left({u (x,y+\Delta y), v (x,y+\Delta y)}\right) - f \left({u (x,y), v (x,y)}\right)} \\
{\widetilde{\Delta }}_{x} f &\equiv & {1 \over 2} {\left\{{f \left({u (x+\Delta x, y), v (x+\Delta x, y)}\right) + f \left({u (x,y), v
(x,y)}\right)}\right\}} \\
{\widetilde{\Delta }}_{y} f &\equiv & {1 \over 2} {\left\{{f \left({u (x,y+\Delta y), v (x,y+\Delta y)}\right) + f \left({u (x,y), v
(x,y)}\right)}\right\}}
\end{eqnarray*}
from which the following identities can be proved:
\begin{eqnarray*}
\Delta f &=& {\Delta }_{u} {\widetilde{\Delta }}_{v} f + {\widetilde{\Delta }}_{u} {\Delta }_{v} f = {\Delta }_{x} {\widetilde{\Delta
}}_{y} f + {\widetilde{\Delta }}_{x} {\Delta }_{y} f \\
\widetilde{\Delta } f &=& {\widetilde{\Delta }}_{u} {\widetilde{\Delta }}_{v} f + {1 \over 4} {\Delta
}_{u} {\Delta }_{v} f = {\widetilde{\Delta }}_{x} {\widetilde{\Delta }}_{y} f + {1 \over 4} {\Delta
}_{x} {\Delta }_{y} f
\end{eqnarray*}

We can define also the difference operators
\begin{eqnarray*}
{{\Delta }_{u_x}} f &\equiv& f (u+{\Delta }_{x}u , v) - f (u,v) \\
{{\Delta }_{u_y}} f &=& f (u+{\Delta }_{y}u , v) - f (u,v) \\
{{\Delta }_{v_x}} f &=& f (u , v +{\Delta }_{x}v) - f (u,v) \\
{{\Delta }_{v_y}} f &=& f (u , v +{\Delta }_{y}v) - f (u,v)
\end{eqnarray*}
and similarly for the promediate operator
\begin{eqnarray*}
{{\widetilde{\Delta }}_{u_x}} f &=& {1 \over 2} \left\{{f (u+{\Delta}_{x}u , v) + f (u,v)}\right\} \\
{{\widetilde{\Delta }}_{u_y}} f &=& {1 \over 2} \left\{{f (u+{\Delta }_{y}u , v) + f (u,v)}\right\} \\
{{\widetilde{\Delta }}_{v_x}} f &=& {1 \over 2} \left\{{f (u , v +{\Delta }_{x}v) + f (u,v)}\right\} \\
{{\widetilde{\Delta }}_{v_y}} f &=& {1 \over 2} \left\{{f (u , v +{\Delta }_{y}v) + f (u,v)}\right\}
\end{eqnarray*}

From which we deduce the following identities:
\begin{eqnarray}
{\Delta }_{x} f &=& {{\Delta }_{u_x}} {{\widetilde{\Delta }}_{v_x}} f + {{\widetilde{\Delta }}_{u_x}}
{{\Delta }_{v_x}} f \\
{\Delta }_{y} f &=& {{\Delta }_{u_y}} {{\widetilde{\Delta }}_{v_y}} f + {{\widetilde{\Delta }}_{u_y}}
{{\Delta }_{v_y}} f \\
{\widetilde{\Delta }}_{x} f &=& {{\widetilde{\Delta }}_{u_x}} {{\widetilde{\Delta }}_{v_x}} f + {1 \over
4}{{\Delta }_{u_x}}\ {{\Delta }_{v_x}} f \\
{\widetilde{\Delta }}_{y} f &=& {{\widetilde{\Delta }}_{u_y}} {{\widetilde{\Delta }}_{v_y}} f + {1 \over
4}{{\Delta }_{u_y}}\ {{\Delta }_{v_y}} f
\end{eqnarray}

These formulas can easily be applied to vector-valued functions:
$$\vec{u} = \left({{u}_{1}(x), {u}_{2}(x), \ldots , {u}_{n}(x)}\right) = \vec{u}(x)$$
and its ``tanget vector''
$${\Delta \vec{ u} \over \Delta x} = \left({{\Delta { u}_{ 1} \over \Delta x}, {\Delta { u}_{ 2}
\over \Delta x}, \ldots , {\Delta { u}_{ n} \over \Delta x}}\right) \equiv \vec{v}(x)$$

An inmediate application is the four-position and four-velocity vectors in special relativity:
\begin{eqnarray*}
{x}^{\mu }(\tau ) &\equiv& \left({{x}^{1} (\tau ), {x}^{2} (\tau ), {x}^{3} (\tau ), {x}^{4} (\tau
)}\right) \\
{V}^{\mu }(\tau ) &\equiv& \left({{{\Delta x}^{1} \over \Delta \tau }, {{\Delta x}^{2} \over \Delta \tau
}, {{\Delta x}^{3} \over \Delta \tau }, {{\Delta x}^{4} \over \Delta \tau } }\right)
\end{eqnarray*}

These vector-valued vector can be expressed as
$$\vec{u}= {u}^{a} {\vec{e}}_{a}$$
for a given set of orthonormal vectors ${\vec{e}}_{a}$

\section{Discrete differential forms}

\setcounter{equation}{0}
Given a vectorial space $V^n$ over $\cal Z$ we can define a real-valued linear function over $\cal Z$
\begin{equation}
f\left({u}\right) \equiv \left\langle{\omega , u}\right\rangle \quad u\in { V}^{ n}
\end{equation}

The forms $\omega$ constitue a vectorial linear space (dual space) ${}^*{V}^{n}$, and can be
expanded in terms of a basis ${\omega }^{\alpha }$
$$\omega = {\sigma }_{\alpha } {\omega }^{\alpha }$$

The basis $e_{\beta}$ of $V^n$ and $\omega^{\alpha}$ of ${}^*{V}^{n}$ can be contracted in the following
way
\begin{equation}
\left\langle{{\omega }^{\alpha } , {e}_{\beta }}\right\rangle {\delta }_{\beta}^{\alpha } 
\end{equation}
hence 
\begin{equation}
\left\langle{\omega , {e}_{a}}\right\rangle = {\sigma }_{\alpha } \quad , \quad 
\left\langle{{\omega }^{\beta } , u}\right\rangle = {u}^{\beta } \quad , \quad \left\langle{\omega ,
u}\right\rangle = {\sigma }_{\alpha } {u}^{\alpha }
\end{equation}

If we take ${\omega }^{\beta } = \Delta {x}^{\beta }$ as coordinate basis for the linear forms we can
construct discrete differential forms (a discrete version of the continuous differential forms)

A particular example of this discrete form is the total difference operator (1,3) of a function
of several discrete variables written in the following way:
\begin{equation}
\Delta f(x,y) = \left({{{\Delta }_{x} {\widetilde{\Delta }}_{y} f \over \Delta
x}}\right) \Delta x +
\left({{{\widetilde{\Delta }}_{x} {\Delta }_{y} f \over \Delta y}}\right) \Delta y
\end{equation}

For these discrete forms we can define the exterior product of two form $\sigma$ and $\delta$
$$\rho \wedge \sigma = -\sigma \wedge \rho $$
which is linear in both arguments.

For the coordinate basis we also have
$$\Delta x \wedge \Delta y = -\Delta y \wedge \Delta x$$

With the help of this exterior product we can construct a second order discrete differential
form or 2-form, namely 
\begin{equation}
\rho \wedge \sigma = -{\rho }_{\alpha } {\Delta x}^{\alpha } \wedge {\sigma }_{\alpha
} {\Delta x}^{\beta } = {1 \over 2} \left({{\rho }_{\alpha } {\sigma }_{\beta } - {\rho }_{\beta
}{\sigma }_{\alpha }}\right) {\Delta x}^{\alpha } \wedge {\Delta x}^{\beta } \equiv {\sigma }_{\alpha
\rho } {\Delta x}^{\alpha } \wedge {\Delta x}^{\beta }
\end{equation}
where ${\sigma }_{\alpha \rho }$ is an antisymmetric tensor. Similarly we can define a discrete
p-form in a $n$-dimensional space $(p<n)$
$$
\sigma = {1 \over P!} {\sigma }_{i_1}{}_{i_2}\ldots {}_{i_p} {\Delta x}^{{i}_{1}}
\wedge {\Delta x}^{{i}_{2}} \ldots \wedge {\Delta x}^{{i}_{p}}
$$
where ${\sigma }_{i_1}{}_{i_2}\ldots {}_{i_p}$ is a totally antysymmetric tensor

The dual of a $p$-form in a $n$-dimensional space is the $(n-p)$-form $^*\alpha$ with components
$$
{\left({{}^*\alpha }\right)}_{{k}_{1} {k}_{2}\ldots {k}_{n-p}} = {1 \over p!} {\alpha }^{{i}_{1}
{i}_{2}\ldots {i}_{p}} {\left.{\varepsilon }\right.}_{{i}_{1} {i}_{2}\ldots {i}_{p}{k}_{1}\ldots
{k}_{n-p}}
$$
where $\varepsilon$ is the $n$-dimensional Levy-Civitt\'a totally antisymmetric tensor
$\left({{\varepsilon }_{1 2 3 \ldots } \equiv 1}\right)$

We give now some examples:

\bigskip
\noindent { \underline {Energy-momentum 1-form}}
\begin{equation}
{\bf P} = -E \Delta t + {P}_{x} \Delta x + {P}_{y} \Delta y + {P}_{z} \Delta z
\end{equation}
where $\left({{P}_{x}, {P}_{y}, {P}_{z}, iE}\right) \equiv {P}_{n}$ is the four-momentum.

\bigskip
\noindent { \underline {Vector potential 1-form}}
\begin{equation}
{\bf A} = {A}_{\mu } {\Delta x}^{\mu } = {A}_{x} \Delta x + {A}_{y} \Delta y + {A}_{z} \Delta z + {A}_{t}
\Delta t
\end{equation}
where $A_{\mu} = \left({{A}_{x}, {A}_{y}, {A}_{z}, {A}_{t}}\right)$ is the four-potential.

\bigskip
\noindent { \underline {Charge-current 1-form}}
\begin{equation}
{\bf J} = {J}_{\mu } {\Delta x}^{\mu } = {J}_{x} \Delta x + {J}_{y} \Delta y +
{J}_{z} \Delta z -
\rho \Delta t
\end{equation}
where ${J}_{\mu }$ is the density current four-vector.

\bigskip
\noindent { \underline {Faraday 2-form}}
\begin{eqnarray}
{\bf F} &=& E_{x} \Delta x \wedge \Delta t + {E}_{y} \Delta y \wedge \Delta t + {E}_{z}\Delta z \wedge \Delta t \nonumber\\ 
&+ &  {B}_{x} \Delta y \wedge \Delta z + {B}_{y} \Delta z \wedge \Delta x + {B}_{z} \Delta x
\wedge \Delta y = {1 \over 2} {F}_{\mu \nu } {\Delta x}^{\mu } \wedge {\Delta x}^{\nu }
\end{eqnarray}
with $\left({{B}_{x} , {B}_{y} , {B}_{z}}\right) \equiv \vec{B}$ and $\left({{E}_{x} , {E}_{y} , {E}_{z}}\right) \equiv \vec{E}$ the
magnetic and electric field, respectively.

\bigskip
\noindent { \underline {Maxwell 2-form (dual of Faraday 2-form)}}
\begin{eqnarray}
{}^*{\bf F} = {1 \over 2} {\varepsilon }_{\mu \nu \lambda \kappa } {F}^{\lambda \kappa }
{\Delta x}^{\mu } \wedge {\Delta x}^{\nu }
 = &-&{B}_{x} \Delta x \wedge \Delta t - {B}_{y} \Delta y \wedge \Delta t - {B}_{z} \Delta z \wedge \Delta t \nonumber  \\
&+& {E}_{x} \Delta y \wedge \Delta z + {E}_{y}
\Delta z \wedge \Delta x + {E}_{z} \Delta x \wedge \Delta y
\end{eqnarray}

\section{Exterior calculus}

\setcounter{equation}{0}

Given a 1-form in a two-dimensional space
$$\omega = a(x,y) \Delta x + b(x,y) \Delta y$$
we can define the exterior difference, in the similar way as the exterior derivative, namely, 
\begin{eqnarray}
\Delta \omega &\equiv& \Delta a \wedge \Delta x + \Delta b \wedge \Delta y \nonumber \\
&=& \left({{{\Delta }_{x}{\widetilde{\Delta }}_{y} a \over \Delta x} \Delta x + {{\widetilde{\Delta }}_{x} {\Delta }_{y} a
\over \Delta y} \Delta y}\right) \wedge \Delta x + \left({{{\Delta }_{x} {\widetilde{\Delta }}_{y} b
\over \Delta x} \Delta x + {{\widetilde{\Delta }}_{x} {\Delta }_{y} b \over \Delta y} \Delta
y}\right) \wedge \Delta y \nonumber \\
&=& \left({{{\Delta }_{x} {\widetilde{\Delta }}_{y} b \over \Delta x} -
{{\widetilde{\Delta }}_{x} {\Delta }_{y} a \over \Delta y}}\right) \Delta x \wedge \Delta y
\end{eqnarray}
where in the last expression we have used the properties of the exterior product.

This definition of exterior difference can be easily written for 1-form in $n$-dimensional space.

Given a 2-form in a 3-dimensional space,
\begin{equation}
\omega = a (x,y,z) \Delta y \wedge \Delta z + b(x,y,z) \Delta z \wedge \Delta x + c
(x,y,z) \Delta x \wedge \Delta y
\end{equation}
we can also define the exterior difference as:
\begin{eqnarray}
\Delta \omega &=& \Delta a \wedge \Delta y \wedge \Delta z + \Delta b \wedge \Delta z
\wedge \Delta x + \Delta c \wedge \Delta x \wedge \Delta y \nonumber \\
&=& \left({{{\Delta }_{x} {\widetilde{\Delta
}}_{y} {\widetilde{\Delta }}_{z} a \over \Delta x} + {{\widetilde{\Delta }}_{x} {\Delta }_{y}
{\widetilde{\Delta }}_{z} b \over \Delta y} + {{\widetilde{\Delta }}_{x} {\widetilde{\Delta }}_{y}
{\Delta }_{z} c \over \Delta z}}\right) \Delta x \wedge \Delta y \wedge \Delta z
\end{eqnarray}

Given a 3-form in a 4-dimensional space
$$
\omega = a \Delta y \wedge \Delta z + \Delta t + b \Delta z \wedge \Delta t \wedge \Delta x + c
\Delta t \wedge \Delta x \wedge \Delta y + d \Delta x \wedge \Delta y \wedge \Delta z
$$
we can define an exterior difference as before:
\begin{eqnarray}
\omega &=& \Delta a \wedge \Delta y \wedge \Delta z \wedge \Delta t + \Delta b \wedge
\Delta z \wedge \Delta t \wedge \Delta x \nonumber \\
& & + \Delta c \wedge \Delta t \wedge \Delta x \wedge \Delta y + \Delta d \wedge
\Delta x \wedge \Delta y \wedge \Delta z \nonumber \\
&=& \left({{\Delta }_{x} {\widetilde{\Delta }}_{y}
{\widetilde{\Delta }}_{z} {\widetilde{\Delta }}_{t} a \over \Delta x} - {{\widetilde{\Delta }}_{x}
{\Delta }_{y} {\widetilde{\Delta }}_{z} {\widetilde{\Delta }}_{t} b \over \Delta y} +
{{\widetilde{\Delta }}_{x} {\widetilde{\Delta }}_{y} {\Delta }_{z} {\widetilde{\Delta }}_{t} c \over
\Delta z}\right. \nonumber \\
& & - \left.{{\widetilde{\Delta }}_{x} {\widetilde{\Delta }}_{y} {\widetilde{\Delta }}_{z} {\Delta
}_{t} d \over \Delta t}\right) \Delta x \wedge \Delta y \wedge \Delta z \wedge \Delta t
\end{eqnarray}

The exterior derivative applied to the product of a 0-form (scalar function f) and a 1-form $(\omega
= a \Delta x + b \Delta y)$ is
\begin{equation}
\Delta \left({f\omega }\right) = \widetilde{\Delta }f \Delta \omega + \Delta f \wedge
\widetilde{\Delta }\omega 
\end{equation}
where $\widetilde{\Delta }f$ is expressed in (1.4) and $\widetilde{\Delta }\omega =
\widetilde{\Delta }a \Delta x + \widetilde{\Delta }b \Delta y$

The exterior difference of the product of two 1-forms is easily obtained
\begin{equation}
\Delta \left\{{{\omega }_{1} \wedge {\omega }_{2}}\right\} = \Delta {\omega }_{1}
\wedge
\widetilde{\Delta } {\omega }_{2} - \widetilde{\Delta } {\omega }_{1} \wedge \Delta {\omega }_{2}
\end{equation}

The exterior difference of the product of a p-form $\rho$ and a q-form $\sigma$ is
\begin{equation}
\Delta \left\{{\rho \wedge \sigma }\right\} = \Delta \rho \wedge \widetilde{\Delta }
\sigma + {\left({-1}\right)}^{p} \widetilde{\Delta } \rho \wedge \Delta \sigma
\end{equation}

Finally for any p-form $\omega$ we have
\begin{equation}
{\Delta }^{2} \omega = \Delta \left({\Delta \omega }\right) = 0
\end{equation}

Some examples:

From the Faraday 2-form we write down one set of Maxwell difference equations
$$
\Delta {\bf F} = \Delta \left({\Delta {\bf A}}\right) = 0
$$
\begin{eqnarray}
\left({{{\Delta }_{x} {\widetilde{\Delta }}_{y} {\widetilde{\Delta }}_{z}
{\widetilde{\Delta }}_{t} {B}_{x} \over \Delta x} + {{\widetilde{\Delta }}_{x} {\Delta }_{y}
{\widetilde{\Delta }}_{z} {\widetilde{\Delta }}_{t} {B}_{y} \over \Delta y} + {{\widetilde{\Delta
}}_{x} {\widetilde{\Delta }}_{y} {\Delta }_{z} {\widetilde{\Delta }}_{t} {B}_{z} \over \Delta
z}}\right) \Delta x \wedge \Delta y \wedge \Delta z \nonumber \\
+ \left({{{\widetilde{\Delta }}_{x}
{\widetilde{\Delta }}_{y} {\widetilde{\Delta }}_{z} {\Delta }_{t} {B}_{x} \over \Delta t} +
{{\widetilde{\Delta }}_{x} {\Delta }_{y} {\widetilde{\Delta }}_{z} {\widetilde{\Delta }}_{t} {E}_{z}
\over \Delta y} - {{\widetilde{\Delta }}_{x} {\widetilde{\Delta }}_{y} {\Delta }_{z}
{\widetilde{\Delta }}_{t} {E}_{y} \over \Delta z}}\right) \Delta t \wedge \Delta y \wedge \Delta z  \nonumber \\
+\left({{{\widetilde{\Delta }}_{x} {\widetilde{\Delta }}_{y} {\widetilde{\Delta }}_{z} {\Delta }_{t}
{B}_{y} \over \Delta t} + {{\widetilde{\Delta }}_{x} {\widetilde{\Delta }}_{y} {\Delta }_{z}
{\widetilde{\Delta }}_{t} {E}_{x}
\over \Delta z} - {{\Delta }_{x} {\widetilde{\Delta }}_{y} {\widetilde{\Delta }}_{z}
{\widetilde{\Delta }}_{t} {E}_{z} \over \Delta x}}\right) \Delta t \wedge \Delta z \wedge \Delta x  \nonumber \\
+\left({{{\widetilde{\Delta }}_{x} {\widetilde{\Delta }}_{y} {\widetilde{\Delta }}_{z} {\Delta }_{t}
{B}_{z} \over \Delta t} + {{\Delta }_{x} {\widetilde{\Delta }}_{y} {\widetilde{\Delta }}_{z}
{\widetilde{\Delta }}_{t} {E}_{y} \over \Delta x} - {{\widetilde{\Delta }}_{x} {\Delta }_{y}
{\widetilde{\Delta }}_{z} {\widetilde{\Delta }}_{t} {E}_{x} \over \Delta y}}\right) \Delta t \wedge
\Delta x \wedge \Delta y
\end{eqnarray}
from the Maxwell 2-dual form ${}^*{\bf F}$ we get the other set of Maxwell equations:
$$
\Delta {}^*{\bf F} = 4\pi {}^*{\bf J}
$$
\begin{eqnarray}
&{}&\left({{{\Delta }_{x} {\widetilde{\Delta }}_{y} {\widetilde{\Delta }}_{z} {\widetilde{\Delta }}_{t}
{E}_{x} \over \Delta x} + {{\widetilde{\Delta }}_{x} {\Delta }_{y} {\widetilde{\Delta }}_{z}
{\widetilde{\Delta }}_{t} {E}_{y} \over \Delta y} + {{\widetilde{\Delta }}_{x} {\widetilde{\Delta
}}_{y} {\Delta }_{z} {\widetilde{\Delta }}_{t} {E}_{z} \over \Delta z}}\right) \Delta x \wedge \Delta
y \wedge \Delta z \nonumber \\ 
&+& \left({{{\widetilde{\Delta }}_{x} {\widetilde{\Delta }}_{y} {\widetilde{\Delta
}}_{z} {\Delta }_{t} {E}_{x} \over \Delta t} - {{\widetilde{\Delta }}_{x} {\Delta }_{y}
{\widetilde{\Delta }}_{z} {\widetilde{\Delta }}_{t} {B}_{z} \over \Delta y} + {{\widetilde{\Delta }}_{x} {\widetilde{\Delta
}}_{y} {\Delta }_{z} {\widetilde{\Delta }}_{t} {B}_{y} \over \Delta z}}\right) \Delta t \wedge \Delta
y \wedge \Delta z \nonumber \\ 
&+& \left({{{\widetilde{\Delta }}_{x} {\widetilde{\Delta }}_{y} {\widetilde{\Delta
}}_{z} {\Delta }_{t} {E}_{y} \over \Delta t} - {{\widetilde{\Delta }}_{x} {\widetilde{\Delta }}_{y}
{\Delta }_{z} {\widetilde{\Delta }}_{t} {B}_{x} \over \Delta z} + {{\Delta }_{x} {\widetilde{\Delta
}}_{y} {\widetilde{\Delta }}_{z} {\widetilde{\Delta }}_{t} {B}_{z} \over \Delta x}}\right) \Delta t
\wedge \Delta z \wedge \Delta x \nonumber \\ 
&{+}& \left({{{\widetilde{\Delta }}_{x} {\widetilde{\Delta }}_{y}
{\widetilde{\Delta }}_{z} {\Delta }_{t} {E}_{z} \over \Delta t} - {{\Delta }_{x} {\widetilde{\Delta
}}_{y} {\widetilde{\Delta }}_{z} {\widetilde{\Delta }}_{t} {B}_{y} \over \Delta x} +
{{\widetilde{\Delta }}_{x} {\Delta }_{y} {\widetilde{\Delta }}_{z} {\widetilde{\Delta }}_{t} {B}_{x}
\over \Delta y}}\right) \Delta t \wedge \Delta x \wedge \Delta y \nonumber \\ 
&=& 4\pi \left({\rho \Delta x \wedge
\Delta y \wedge \Delta z -{J}_{x} \Delta t \wedge \Delta y \wedge \Delta z} \right.\nonumber \\
& &- \left. {{J}_{y} \Delta t \wedge
\Delta z \wedge \Delta x - {J}_{z} \Delta t \wedge \Delta x \wedge \Delta y}\right)
\end{eqnarray}

Taking the exterior derivative of the last equation we get an other example of ${\Delta }^{2} =0$.
\begin{equation}
\left({{{\widetilde{\Delta }}_{x} {\widetilde{\Delta }}_{y} {\widetilde{\Delta }}_{z}
{\Delta }_{t}
\rho \over \Delta t} + {{\Delta }_{x} {\widetilde{\Delta }}_{y} {\widetilde{\Delta }}_{z}
{\widetilde{\Delta }}_{t} {J}_{x} \over \Delta x} + {{\widetilde{\Delta }}_{x} {\Delta }_{y}
{\widetilde{\Delta }}_{z} {\widetilde{\Delta }}_{t} {J}_{y} \over \Delta y} + {{\widetilde{\Delta
}}_{x} {\widetilde{\Delta }}_{y} {\Delta }_{z} {\Delta }_{t} {J}_{z} \over \Delta z}}\right) \cdot
\Delta t \wedge \Delta x \wedge \Delta y \wedge \Delta z = 0
\end{equation}

Note that the coefficient of the difference form is the discrete version of the continuity equation.

From a scalar function we get the wave equations in terms of difference operators, namely, 
\begin{equation}
-{}^*\Delta {}^*\Delta \phi \equiv \raisebox{0,5ex}{\fbox{\rule{0mm}{0,5ex}\hspace*{0,5ex}}}\; \phi 
\end{equation}
where \raisebox{0,5ex}{\fbox{\rule{0mm}{0,5ex}\hspace*{0,5ex}}} is the discrete d'Alambertian operator:
\begin{eqnarray}
\left\{ - {\widetilde{\nabla }}_{x}{\widetilde{\nabla }}_{y}{\widetilde{\nabla
}}_{z}{\nabla }_{t}\left({{\widetilde{\Delta }}_{x}{\widetilde{\Delta }}_{y}{\widetilde{\Delta
}}_{z}{\Delta }_{t}}\right)+{\nabla }_{x}{\widetilde{\nabla }}_{y}{\widetilde{\nabla
}}_{z}{\widetilde{\nabla }}_{t}\left({{\Delta }_{x}{\widetilde{\Delta }}_{y}{\widetilde{\Delta
}}_{z}{\widetilde{\Delta }}_{t}}\right) \right. \nonumber \\
\left. +{\widetilde{\nabla }}_{x}{\nabla }_{y}{\widetilde{\nabla
}}_{z}{\widetilde{\nabla }}_{t}\left({{\widetilde{\Delta }}_{x}{\Delta }_{y}{\widetilde{\Delta
}}_{z}{\widetilde{\Delta }}_{t}}\right)+{\widetilde{\nabla }}_{x}{\widetilde{\nabla }}_{y}{\nabla
}_{z}{\widetilde{\nabla }}_{t}\left({{\widetilde{\Delta }}_{x}{\widetilde{\Delta }}_{y}{\Delta
}_{z}{\widetilde{\Delta }}_{t}}\right)\right\}\phi \left({ xyzt}\right) =0
\end{eqnarray}

From the vector potential 1-forms ${\bf A} = A_\mu \Delta x^\mu$ we can construct Faraday 2-form
${\bf F} =  \Delta {\bf A}$ from which the Maxwell equations are derived
$$
\Delta {}^*{\bf F}=\Delta {}^*\Delta {\bf A}=4\pi {}^*{\bf J}
$$

Taking the dual of this expression we obtain
\begin{equation}
{}^*\Delta {}^*\Delta A=4\pi J
\end{equation}

If we choose the Lorentz condition
\begin{equation}
{{\Delta }_{x}{\widetilde{\Delta }}_{y}{\widetilde{\Delta }}_{z}{\widetilde{\Delta
}}_{t}{A}_{x} \over \Delta x}+{{\widetilde{\Delta }}_{x}{\Delta }_{y}{\widetilde{\Delta
}}_{z}{\widetilde{\Delta }}_{t}{A}_{y} \over \Delta y}+{{\widetilde{\Delta
}}_{x}{\widetilde{\Delta }}_{y}{\Delta }_{z}{\widetilde{\Delta }}_{t} {A}_{z} \over \Delta z}-{{\widetilde{\Delta }}_{x} 
{\widetilde{\Delta }}_{y}{\widetilde{\Delta }}_{z}{\Delta }_{t}{A}_{t} \over \Delta t}=0
\end{equation}
we obtain the wave equation for the vector potential
\begin{equation}
\raisebox{0,5ex}{\fbox{\rule{0mm}{0,5ex}\hspace*{0,5ex}}}\; {A}_{\mu }=4\pi { J}_{\mu }
\end{equation}
where \raisebox{0,5ex}{\fbox{\rule{0mm}{0,5ex}\hspace*{0,5ex}}} is the d'Alambertian defined in
(3.13)

\section{Lorentz transformations}

\setcounter{equation}{0}

In order to compute the transformation of the discrete differential forms we start with the
coordinate-independent nature of 1-form

\begin{equation}
\omega ={\omega }_{\mu }\Delta { x}^{\mu }
\end{equation}
where the $\Delta x^{\mu }$ are the space-time intervals in Minskowski space-time. From

\begin{equation}
\Delta { x}^{\mu \prime} ={\Lambda }_{\nu }^{\mu \prime}\Delta { x}^{\nu }
\end{equation}
where ${\Lambda }_{\nu }^{\mu \prime}$ is a global Lorentz transformation, and from the
coordinate-free expresion for $\omega$ we get

\begin{equation}
{\omega }_{\mu \prime}={\omega }_{\nu }{\Lambda }_{\mu \prime}^{\nu }
\end{equation}
Recall that ${\Lambda }_{\mu \prime}^{\nu }{\Lambda }_{\rho }^{\mu \prime}={\delta }_{\rho }^{\nu }$

From the total difference of a function of several variables $f\left({x,y,z,t}\right)$

\begin{equation}
\Delta f={{\Delta }_{x}{\widetilde{\Delta }}_{y}{\widetilde{\Delta
}}_{z}{\widetilde{\Delta }}_{t}f \over \Delta x}\Delta x+{{\widetilde{\Delta }}_{x}{\Delta
}_{y}{\widetilde{\Delta }}_{z}{\widetilde{\Delta }}_{t}f \over \Delta y}\Delta y+{{\widetilde{\Delta
}}_{x}{\widetilde{\Delta }}_{y}{\Delta }_{z}{\widetilde{\Delta }}_{t}f \over \Delta z}\Delta
z+{{\widetilde{\Delta }}_{x}{\widetilde{\Delta }}_{y}{\widetilde{\Delta }}_{z}{\Delta }_{t}f \over
\Delta t}\Delta t
\end{equation}
it follows that the coefficients of the 1-forms, namely,

\begin{equation}
\left({{{\Delta }_{x}{\widetilde{\Delta }}_{y}{\widetilde{\Delta
}}_{z}{\widetilde{\Delta }}_{t}f
\over \Delta x},{{\widetilde{\Delta }}_{x}{\Delta }_{y}{\widetilde{\Delta }}_{z}{\widetilde{\Delta
}}_{t}f \over \Delta y},{{\widetilde{\Delta }}_{x}{\widetilde{\Delta }}_{y}{\Delta
}_{z}{\widetilde{\Delta }}_{t} \over \Delta z},{{\widetilde{\Delta }}_{x}{\widetilde{\Delta
}}_{y}{\widetilde{\Delta }}_{z}{\Delta }_{t} \over \Delta t}}\right)
\end{equation}
transform covariantly like the coefficients ${\omega }_{\mu }$ of (4.1). Note that in this case the
meaning of $\Delta x,\Delta y,\Delta z,\Delta t$ in the denominator is
different from the meaning in the numerator, because the later $\Delta { x}^{\mu }$ are elements
of the exterior products, and the former $\Delta { x}$ are small scalars. The different roll of these quantities becomes clear in the
continuous limit, where 
$\Delta x\rightarrow ds$ and ${\Delta x\widetilde{\Delta }y\widetilde{\Delta }z\Delta t\over \Delta
x}\rightarrow {\partial \over \partial x}$

For the 2-form in Minkowski space

$${\bf F}={1 \over 2}{F}_{\mu \nu }\Delta { x}^{\mu } \wedge \Delta { x}^{\nu }$$
 we obtain

$${\bf F}={1 \over 2}{F}_{\mu \prime\nu \prime}\Delta { x}^{\mu \prime} \wedge \Delta { x}^{\nu
\prime}$$
because of the coordinate independent nature of the Faraday 2-form. From the transformation
of $\Delta { x}^{\mu}$ (see (4.2)) and the properties of the exterior product we get

\begin{equation}
{F}_{\mu \prime\nu \prime}={F}_{\alpha \beta }{\Lambda }_{\mu \prime}^{\alpha
}{\Lambda }_{\nu
\prime}^{\beta }
\end{equation}

The same technic can be applied to components of discrete p-forms. The dual of a p-form are also
discrete $(n-p)$-form, therefore their components transform covariantly like totally antisymmetric
tensor. For instance the components of Maxwell 2-forms.

$${\left({{}^*F}\right)}_{\alpha \beta }={\epsilon }_{\alpha \beta \mu \nu }{F}^{\mu \nu }\
\left({{\epsilon }_{1234}=+1}\right)$$
transform covariantly

\begin{equation}
{\left({{}^*F}\right)}_{\alpha \prime\beta \prime}={\left({{}^*F}\right)}_{\kappa
\lambda }{\Lambda }_{\alpha \prime}^{\kappa }{\Lambda }_{\beta \prime}^{\lambda }
\end{equation}

With this definition of the transformation of the components of a discrete p-form we can prove the
covariance of the discrete wave equation, and the covariance of Maxwell equation in discrete form

\section{Application to Klein-Gordon and Dirac wave \newline equation on the lattice}

\setcounter{equation}{0}

We define the scalar function on the (3+1) dimensional cubic lattice

$$\phi \left({{j}_{1}{\epsilon }_{1},{j}_{2}{\epsilon }_{2},{j}_{3}{\epsilon
}_{3},n\tau }\right)\equiv \phi \left({\vec {\jmath},n}\right)$$
where ${\epsilon }_{1},{\epsilon }_{2},{\epsilon }_{3},\tau $ are small quantities in the
space-time directions and ${j}_{1},{j}_{2},{j}_{3},n$ are integer numbers.

We define the difference operators

\begin{eqnarray*}
{\delta }_{\mu }^{+} &\equiv& {1 \over {\epsilon }_{\mu }}{\Delta }_{\mu }\prod_{\nu
\ne \mu } {\widetilde{\Delta }}_{\nu } , \mu ,\nu =1,2,3,4 \\
{\delta }_{\mu }^{-} &\equiv& {1 \over {\epsilon }_{\mu }}{\nabla }_{\mu }\prod_{\nu
\ne \mu } {\widetilde{\nabla }}_{\nu } \\
{\eta }^{+} &\equiv& \prod_{\mu =1}^{ 4} {\widetilde{\Delta }}_{\mu } \\
{\eta }^{-} &\equiv& \prod_{\mu =1}^{ 4} {\widetilde{\nabla }}_{\mu }
\end{eqnarray*}

Then the Klein-Gordon wave equations defined on the grid points of the lattice can be read off

\begin{equation}
\left({{\delta }_{1}^{+}{\delta }_{1}^{-}+{\delta }_{2}^{+}{\delta }_{2}^{-}+{\delta
}_{3}^{+}{\delta }_{3}^{-}-{\delta }_{4}^{+}{\delta }_{4}^{-}-{M}^{2}{\eta }^{+}{\eta
}^{-}}\right)\phi \left({\vec {\jmath},n}\right) =0
\end{equation}

It can be verified by direct substitution that the plane wave solution satisfy the difference
equation

\begin{equation}
f\left({\vec {\jmath},n}\right)\equiv {\left({{1+{1 \over 2}i{\epsilon }_{1}{k}_{1} \over 1-{1
\over 2}i{\epsilon }_{1}{k}_{1}}}\right)}^{{j}_{1}}{\left({{1+{1 \over 2}i{\epsilon }_{2}{k}_{2}
\over 1-{1 \over 2}i{\epsilon }_{2}{k}_{2}}}\right)}^{{j}_{2}}{\left({{1+{1 \over
2}i{\epsilon }_{3}{k}_{3} \over 1-{1 \over 2}i{\epsilon }_{3}{k}_{3}}}\right)}^{{j}_{3}}\
{\left({{1-{1 \over 2}i\tau \omega \over 1+{1 \over 2}i\tau \omega }}\right)}^{n}
\end{equation}
provided the ``dispersion relation'' is satisfied

\begin{equation}
{\omega }^{2}-{k}_{1}^{2}-{k}_{2}^{2}-{k}_{3}^{2}={M}^{2}
\end{equation}

From the last section the Klein-Gordon equation is invariant under finite Lorentz transformations.
Imposing boundary conditions on the plan waves we can construct a complete set of orthogonal
functions, hence a Fourier analysis can be developped as it has been done in Ref. [1].

The discrete version of the Dirac wave equation can be written as

\begin{equation}
\left({{\gamma }_{1}{\delta }_{1}^{+}+{\gamma }_{2}{\delta }_{2}^{+}+{\gamma
}_{3}{\delta }_{3}^{+}-{i\gamma }_{4}{\delta }_{4}^{+}+M{\eta }^{+}}\right)\psi \left({\vec {\jmath},n}\right) =0
\end{equation}
where ${\gamma }_{\mu },\quad \mu =1,2,3,4$ are the usual Dirac matrices. Applying the
operator 

$$\left({{\gamma }_{1}{\delta }_{1}^{-}+{\gamma }_{2}{\delta }_{2}^{-}+{\gamma }_{3}{\delta
}_{3}^{-}-{i\gamma }_{4}{\delta }_{4}^{-}-M{\eta }^{-}}\right)$$
from the left on both sides of (5.4) we recover the Klein-Gordon equation (5.1). Let now construct
solutions to (5.4) of the form

$$\psi \left({\vec {\jmath},n}\right) =\omega \left({\vec{ k} ,E}\right) \
f\left({\vec {\jmath},n}\right)$$
where the $f\left({\vec {\jmath},n}\right)$ are given in (5.2).

The four-component spinors $\omega \left({\vec{ k} ,E}\right)$, with momentum $\vec{k}\
\equiv \left({{k}_{1},{k}_{2},{k}_{3}}\right)$, must satisfy

\begin{equation}
\left({i\vec{\gamma }\cdot \vec{ k} -{\gamma }_{4}E+M}\right)\omega \left({\vec{ k}
,E}\right) =0
\end{equation}
as in the continuous case. Multiplying this equation from the left by

$$\left({i\vec{\gamma }\cdot \vec{ k} -{\gamma }_{4}E-M}\right)$$

we obtain the dispersion relation

\begin{equation}
{E}^{2}-{\vec{k}}^{2}={M}^{2}
\end{equation}

From (5.5) we obtain the spinors solutions corresponding to positive energy as in the continuous case

\begin{equation}
{n}_{k\sigma }\equiv {\omega }_{\sigma }\left({\overline{k},E}\right)={{{E}_{k}+M
\over 2M}}^{1/2}\left({\matrix{\xi_\sigma \cr {{\vec{\sigma }\cdot \vec{ k} \over { E}_{ k} +M}{\xi }_{\sigma
}}\cr}}\right)
\end{equation}

with ${E}_{k}\equiv +\sqrt {{\vec{k}}^{2}+{M}^{2}},\sigma =1,2$ and 
$${\xi
}_{1}=\left({\matrix{1\cr 0\cr}}\right),\quad {\xi }_{2}=\left({\matrix{0\cr
1\cr}}\right)$$

Similarly for negative energy spinors. 

In order to analyze the Lorentz invariant of the Dirac wave equation on the lattice we take the
Lorentz transformations of the difference operators as in (4.3)

\begin{eqnarray}
{\delta }_{\mu }^{+\prime} &=& {\Lambda }_{\mu \nu }{\delta }_{\nu }^{+} \\
{\delta }_{\mu }^{-\prime} &=& {\Lambda }_{\mu \nu }{\delta }_{\nu }^{-}\cdot 
\end{eqnarray}

The linear transformation law for $\psi $

$$\psi \prime =S \psi $$
is determined by the requirement that $\psi \prime$ satisfy the same equation in the transformed
frame as does $\psi $ in the original frame

\begin{equation}
\left({{\gamma }_{\mu }{\delta }_{\mu }^{+\prime}+M}\right)\psi \prime =0
\end{equation}

With (5.8) we have

\begin{equation}
\left({{\Lambda }_{\mu \nu }{\gamma }_{\nu }{\delta }_{\nu }^{+}+M}\right)S \psi =0
\end{equation}

Multiplying from the left by ${S}^{-1}$ we recover the Dirac equation in the original form provided

\begin{equation}{S}^{-1}{\gamma }_{\mu }S={\Lambda }_{\mu \nu }{\gamma }_{\nu }
\end{equation}

A particular solution for the Lorentz transformations of the spinors in the case of rotations is

\begin{equation}
\psi \prime =S\psi =\left({\cos\alpha +i\sin\alpha {\gamma }_{5}}\right)\psi
\end{equation}
where $\alpha$ is the angle of rotation and ${\gamma }_{5}\equiv {\gamma }_{ 1}{\gamma }_{ 2}{\gamma
}_{ 3}{\gamma }_{ 4}$. In the case of pure Lorentz transformation we have

\begin{equation}
\psi \prime =S\psi =ch\beta \psi +sh\beta {\gamma }_{ 4}{\gamma }_{ 2} \overline{\psi}
\end{equation}
with $\overline{\psi }={\gamma }_{4}\psi {}^*$ and $\beta$ is the usual parameter such that $th\beta
={v \over C}$


\begin{thebibliography}{9}

\bibitem{} M. Lorente, ``A New Scheme for the Klein-Gordon and Dirac Fields on the Lattice with Axial
Anomaly''. {\it J. Group Theory in Physics}, {\bf 1}, 105-121 (1993).

\bibitem{} N.S. Manton, ``Connections on Discrete Fibre Bundles'', {\it Comm. Math. Phys.}, {\bf 113},
341-351 (1987).

\bibitem{} N.A. Salingaros, G.P. Wene, ``The Clifford Algebra of Differential Forms'' {\it Acta
Applicandae Mathematicae}, {\bf 4}, 271-292 (1985).

\bibitem{} A. Dimakis, F. M\"uller-Hoissen, ``Discrete Differential Calculus'' {\it J. Math. Phys.}, {\bf
35}, 6703-35 (1994).

\bibitem{} P. Becher, H. Joos, ``The Dirac-K\"ahler equation and Fermions on the lattice'' {\it Z.
Phys.}, {\bf C 15}, 343 (1982).

\bibitem{} Y. Smirnov, A. Turbiner, ``Lie Algebraic discretization of differential equations'' (preprint)
IFUNAM FT 95-68 (1995).

\end{thebibliography}
\end{document}